\documentclass[pre,aps,superscriptaddress,amsmath,amssymb,twocolumn]{revtex4-2}

\usepackage{graphicx}
\usepackage{epstopdf, epsfig}
\usepackage{bm}
\usepackage{amsmath}
\usepackage{amssymb}
\usepackage{subfigure}
\usepackage{caption}
\usepackage{color}
\usepackage{verbatim}
\usepackage{float}
\usepackage[dvipsnames]{xcolor}

\newcommand{\M}{\mathcal{M}}
\newcommand{\PP}{\mathcal{P}}

\newcommand{\overbar}[1]{\mkern 1.5mu\overline{\mkern-1.5mu#1\mkern-1.5mu}\mkern 1.5mu}

\begin{document}

\title{Extended Lattice Boltzmann Model} 

\author{M. H. Saadat}
\author{B. Dorschner}
\author{I. V. Karlin}\thanks{Corresponding author}
\email{ikarlin@ethz.ch}
\affiliation{Department of Mechanical and Process Engineering, ETH Zurich, 8092 Zurich, Switzerland}
\date{\today}

\begin{abstract}
Conventional lattice Boltzmann models for the simulation of fluid dynamics are restricted by an error in the stress tensor that is negligible only for vanishing flow velocity and at a singular value of the temperature.  
To that end, we propose a unified formulation that restores Galilean invariance and isotropy of the stress tensor by introducing an extended equilibrium.
This modification extends lattice Boltzmann models to simulations with higher values of the flow velocity and can be used at temperatures that are higher than the lattice reference temperature, which enhances computational efficiency by decreasing the number of required time steps. 
Furthermore, the extended model remains valid also for stretched lattices, which are useful when flow gradients are predominant in one direction.
The model is validated by simulations of two- and three-dimensional benchmark problems, including the double shear layer flow, the decay of homogeneous isotropic turbulence, the laminar boundary layer over a flat plate and the turbulent channel flow.
\end{abstract}

\maketitle

\section{Introduction}
The lattice Boltzmann method (LBM) solves a Boltzmann-type kinetic equation on a discrete velocity set, forming the links of a space-filling lattice. Efficiency of the LBM makes it attractive for the simulation of a wide range of problems in fluid dynamics, see, e.\ g.,\ \cite{succi,krueger}.

In this paper, we revisit the restrictions of LBM due to the geometry of the discrete velocities. 
It is well known that standard LBM velocities yield a persistent error in the fluid stress tensor, which breaks Galilean invariance and limits the operation range of LBM to a vanishing flow velocity and a singular value of the lattice reference temperature.
Only under these conditions, the error can be ignored. 
While one can cope with this error in most incompressible flow applications \cite{succi,krueger}, it certainly affects high-speed compressible flows \cite{prasianakis2007lattice,prasianakis2008lattice,prasianakis2009,li2012coupling,saadat2019lattice,guo2020efficient,sawant2021consistent} and sometimes even low-speed isothermal cases \citep{clausen2009galilean}.
Moreover, the same error is amplified when using stretched (rectangular) lattices instead of the conventional (cubic) lattice, where in addition to the corrupted Galilean invariance, the stress tensor becomes anisotropic \citep{bouzidi2001lattice,chikatamarla2011comment}.

The extension of LBM beyond its classical operation domain was so far addressed with different techniques, depending on the desired outcome. 
For instance, using standard cubic lattices, flow velocity and temperature range can be extended by adding correction terms to the original LBM \cite{prasianakis2007lattice,prasianakis2008lattice,prasianakis2009,li2012coupling,saadat2019lattice,guo2020efficient,sawant2021consistent}. 
The realization varies among different authors and neither address the general case of rectangular lattices. On the other hand, rectangular lattices may improve the computational efficiency of the LBM by using a coarser mesh in the direction of smaller gradients in the flow. Unlike other approaches of handling non-uniform grids (e.g. Eulerian \citep{patil2009finite,hejranfar2020high} and semi-Lagrangian \citep{kramer2017semi, di2018simulation, saadat2020semi, wilde2020semi} off-lattice LBM or grid refinement techniques \citep{lagrava2012advances,dorschner2016grid}), stretched lattices do not require a substantial change in the standard LBM algorithm.  
Recent works on the stretched LBM restore isotropy of the stress tensor by using multi-relaxation time LBM models \citep{zong2016designing,peng2016hydrodynamically,zecevic2020rectangular}. However, these approaches do not address the flow velocity and temperature restrictions.

In this paper, we propose a unified view on the three aspects of the problem, the velocity range, the temperature range and grid stretching, which all
stem from the same error, induced by constraints of the discrete velocity set.
In particular, we propose to use an extended equilibrium, which restores Galilean invariance and isotropy of the stress tensor, enabling simulations at higher flow velocities, higher temperatures using both cubic and stretched lattices, yielding increased accuracy and efficiency.
 
The paper is organized as follows: 
In section \ref{Sec:DKE}, we start with presenting the discrete kinetic equations, following the standard single-relaxation time lattice Bhatnagar--Gross--Krook (LBGK) setting, as well as the equilibrium and extended equilibrium formulation, followed by the derivation of the model's hydrodynamic limit.
Subsequently, in section \ref{Sec:NUMRES}, we assess validity, accuracy and performance of our model using both two- and three-dimensional 
benchmark problems.
As a first step, we verify Galilean invariance, temperature independence and isotropy of the model on the example of an advected decaying shear wave.
It is shown that the theoretical viscosity is recovered for both cubic as well as stretched lattices in a large range of temperatures and advection velocities.
This also indicates that the model can readily be extended to high-speed compressible flows, provided that it is augmented with a suitable solver for the total energy. 
Next, on the example of homogeneous isotropic turbulence, we demonstrate that a speed-up can be achieved by using an
operating temperature, which is larger than the lattice reference temperature. 
The present model can also be viewed as an alternative to preconditioned LBM \citep{guo2004preconditioned} for accelerating the convergence rate but without the restriction to steady flows.
Finally, accuracy and performance are assessed for rectangular lattices using the doubly periodic shear flow, laminar flow over a flat plate and turbulent  channel flow as examples. 
Conclusions are drawn in Sec.\ \ref{Sec:Conclusion}.

\section{Discrete kinetic equations} \label{Sec:DKE}

\subsection{LBGK}
\label{sec:LBGK}
We consider the LBGK equation for the populations $f_i$, associated to the discrete velocities $\bm{v}_i$ for $i=0,\dots,Q-1$,
\begin{align}
f_i(\bm{x}+\bm{v}_i\delta t,t+\delta t) -f_i(\bm{x},t) =  \omega (f_i^{*} -f_i). \label{eq:LBGK}
\end{align}
The extended equilibrium $f_i^*$, which will be specified below, satisfies the local conservation laws for the density $\rho$ and momentum $\rho \bm{u}$,
\begin{align}
	\rho        &= \sum_{i=0}^{Q-1} f_i^* = \sum_{i=0}^{Q-1} f_i, \label{eq:density} \\
	\rho \bm{u} &= \sum_{i=0}^{Q-1} \bm{v}_i f_i^* = \sum_{i=0}^{Q-1} \bm{v}_i f_i. \label{eq:momentum}
\end{align}
The relaxation parameter $\omega$ is related to the kinematic viscosity $\nu$ as will be shown below,
\begin{align}
	\nu = \left( \frac{1}{\omega} - \frac{1}{2} \right)RT \delta t,\label{eq:nu}
\end{align}
where $T$ is the temperature and $R$ is the gas constant. 
We now proceed with identifying the extended equilibrium.

\subsection{Discrete velocities and factorization}
We use the $D3Q27$ lattice, where $D=3$ denotes the spatial dimension and $Q=27$ is the number of discrete speeds, which are given by
\begin{equation}
	\bm{c}_i=(c_{ix},c_{iy},c_{iz}),\ c_{i\alpha}\in\{-1,0,1\}. 
	\label{eq:d3q27vel}
\end{equation}
With (\ref{eq:d3q27vel}), we define the particles' velocities $\bm{v}_i$ in a stretched cell as
\begin{equation}
\bm{v}_i=(\lambda_{x} c_{ix},\lambda_{y}c_{iy},\lambda_{z} c_{iz}), \label{eq:d3q27velSTRETCHED}
\end{equation}
where $\lambda_{\alpha}$ is the stretching factor in the direction $\alpha$.

The (normalized, $\M_{000}=1$) moments $\M_{lmn}$ are defined using the convention
	\begin{equation}
		l\to x,\ m\to y, n\to z;\ l,m,n=0,1,2,\dots,
	\end{equation}
	and thus
	\begin{equation}
		\rho\M_{lmn}=\sum_{i=0}^{Q-1}{v_{ix}^l v_{iy}^m v_{iz}^n}  f_i,
		\label{eq:mom27}
	\end{equation}
For convenience, we use a more specific notation for the first-order and the diagonal second-order moments,
	\begin{align}
		& \M_{100}=u_x,\     \M_{010}=u_y,\    \M_{001}=u_z, \label{eq:Mu}\\
		& \M_{200}=\PP_{xx},\    \M_{020}=\PP_{yy},\    \M_{002}=\PP_{zz}.\label{eq:MP}
	\end{align}
We essentially follow \cite{karlin2010factorization} and consider a class of factorized populations.
To that end, we define a triplet of functions in the three variables, $u$, $\PP$ and $\lambda$,
\begin{align}
	\Psi_{0}(u,\PP,\lambda) &= 1 - \frac{\PP}{\lambda^2}, 
	\label{eqn:phi0S}
	\\
	\Psi_{1}(u, \PP, \lambda) &= \frac{1}{2}\left(\frac{u}{\lambda}+ \frac{\PP}{\lambda^2}\right),
	\label{eqn:phiPlusS}
	\\
	\Psi_{-1}(u,\PP,\lambda) &= \frac{1}{2}\left(-\frac{u}{\lambda} + \frac{\PP}{\lambda^2}\right).
	\label{eqn:phiMinusS}
\end{align}
For the vectors $\bm{u}$, $\bm{\PP}$, and $\bm{\lambda}$,
	\begin{align}\bm{u}&=(u_x,u_y,u_z),\label{eq:us}\\
		\bm{\PP}&=(\PP_{xx},\PP_{yy},\PP_{zz}),\label{eq:Ps}\\
		\bm{\lambda}&=(\lambda_x,\lambda_y,\lambda_z),\label{eq:lambdas}
	\end{align}
	we consider a product-form associated with the discrete velocities $\bm{v}_i$ (\ref{eq:d3q27velSTRETCHED}),
	\begin{equation}\label{eq:prod}
		\Psi_i(\bm{u},\bm{\PP},\bm{\lambda})= \prod_{\alpha=x,y,z} \Psi_{c_{i\alpha}}(u_\alpha,\PP_{\alpha\alpha},\lambda_\alpha).
	\end{equation}
The normalized moments of the product-form (\ref{eq:prod}),
	\begin{equation}
		\label{eq:mom27UniQuE}
		\M_{lmn}=\sum_{i=0}^{Q-1}v_{ix}^l v_{iy}^m v_{iz}^n \Psi_i,
	\end{equation}
are readily computed thanks to the factorization,
\begin{equation}
	\M_{lmn}=\M_{l00}\M_{0m0}\M_{00n},
\end{equation}
where
	\begin{align}
		\M_{000}&=1,\\
		\M_{l00}&=\left\{\begin{aligned}
			&	\lambda_{x}^{l-1}u_{x},\ & l\ \text{odd}\\
			&	\lambda_{x}^{l-2}\PP_{xx},\ & l\ \text{even}
		\end{aligned}
		\right.,
		\\
		\M_{0m0}&=\left\{\begin{aligned}
			&	\lambda_{y}^{l-1}u_{y},\ & l\ \text{odd}\\
			&	\lambda_{y}^{l-2}\PP_{yy},\ & l\ \text{even}
		\end{aligned}
		\right.,
		\\
		\M_{00n}&=\left\{\begin{aligned}
			&	\lambda_{z}^{l-1}u_{z},\ & l\ \text{odd}\\
			&	\lambda_{z}^{l-2}\PP_{zz},\ & l\ \text{even}
		\end{aligned}
		\right..
	\end{align}
For any stretching (\ref{eq:lambdas}), the six-parametric family of normalized populations (\ref{eq:prod}) is identified by the flow velocity (\ref{eq:us}) and the diagonal of the pressure tensor at unit density (\ref{eq:Ps}), and was termed the unidirectional quasi-equilibrium in Ref.\ \cite{karlin2010factorization}. We make use of the product-form (\ref{eq:prod}) to construct all pertinent populations, the equilibrium and the extended equilibrium.

\subsection{Equilibrium and extended equilibrium}

The equilibrium $f_i^{\rm eq}$ is defined by setting $\PP_{\alpha\alpha}$ (\ref{eq:MP}) equal to the equilibrium diagonal element of the pressure tensor at unit density,
\begin{align}
    \PP_{\alpha\alpha}^{\rm eq}=RT+u_{\alpha}^2. \label{eqn:Peqa}
\end{align}
Substituting (\ref{eqn:Peqa}) into (\ref{eq:prod}), we get
\begin{align}
	f_{i}^{\rm eq}&= \rho \Psi_i(\bm{u},\bm{\PP}^{\rm eq},\bm{\lambda}).
\label{eq:27eq}
\end{align}
With (\ref{eq:mom27UniQuE}), we find the pressure tensor and the third-order moment tensor at the equilibrium (\ref{eq:27eq}) as follows,
	\begin{align}
	\bm{P}^{\rm eq} &= \sum\limits_{i = 0}^{Q-1} \bm{v}_{i}\otimes\bm{v}_{i}f_i^{\rm eq} = \bm{P}^{\rm MB}, \label{eqn:Peq}\\
	\bm{Q}^{\rm eq} &=\sum\limits_{i = 0}^{Q-1} \bm{v}_{i}\otimes\bm{v}_{i}\otimes\bm{v}_{i}f_i^{\rm eq}=\bm{Q}^{\rm MB}+\tilde{\bm{Q}}. \label{eqn:Qeq}
\end{align}
The isotropic parts, $\bm{P}^{\rm MB}$ and $\bm{Q}^{\rm MB}$, are the Maxwell--Boltzmann (MB) pressure tensor and the third-order moment tensor, respectively,
\begin{align}
	\bm{P}^{\rm MB}&=p\bm{I}+\rho \bm{u}\otimes\bm{u},\label{eqn:PMB}\\
		\bm{Q}^{\rm MB}&=  {\rm sym}(p\bm{I}\otimes\bm{u})+\rho \bm{u}\otimes\bm{u}\otimes\bm{u},
		\label{eqn:QMB} 
\end{align}
where $p=\rho RT$ is the pressure, ${\rm sym}(\dots)$ denotes symmetrization and $\bm{I}$ is the unit tensor.
The anisotropy of the equilibrium (\ref{eq:27eq}) manifests with the deviation $\tilde{\bm{Q}} =\bm{Q}^{\rm eq}- \bm{Q}^{\rm MB}$ in (\ref{eqn:QMB}), where only the diagonal elements are non-vanishing,
\begin{align}
	\tilde{{Q}}_{\alpha\beta\gamma}= \left\{\begin{aligned}
		&\rho u_{\alpha}(\lambda_{\alpha}^2-3 RT)-\rho u_{\alpha}^3, &\text{ if }\alpha = \beta = \gamma, & \\ 
		&0, & \text{otherwise}.&\\
	\end{aligned}\right. 
\label{eqn:Deviation}
\end{align}
The origin of the {\it diagonal anomaly} (\ref{eqn:Deviation}) is the geometric constraint, $v_{i\alpha}^3 = \lambda_{\alpha}^2 v_{i\alpha}$, which is imposed by the choice of the discrete speeds (\ref{eq:d3q27vel}), and is well known in the case of the standard (unstretched) lattice with $\lambda_{\alpha}=1$.
A remedy in the latter case is to minimize spurious effects of anisotropy by fixing the temperature $T=T_L$,
	\begin{equation} 
		T_L=\frac{1}{3R},\label{eq:TL}
	\end{equation} 
in order to eliminate the linear term $\sim u_{\alpha}$ in  (\ref{eqn:Deviation}). Thus, the use of the equilibrium (\ref{eq:27eq}) in the LBGK equation (\ref{eq:LBGK}) imposes a two-fold restriction on the operation domain: the temperature cannot be chosen differently from (\ref{eq:TL}) while at the same time the flow velocity has to be maintained asymptotically vanishing. Moreover, for stretched lattices, the anisotropy becomes even more pronounced since it is impossible to eliminate the linear deviation in all directions simultaneously by fixing any temperature.

Alternatively, the spurious anisotropy effects can be canceled out by extending the equilibrium 
such that the third-order moment anomaly is compensated in the hydrodynamic limit.
Because the anomaly only concerns the diagonal (unidirectional) elements of the third-order moments, the cancellation can be achieved by redefining the diagonal elements of the second-order moments. 
As demonstrated below, in order the achieve cancellation of the errors,  
the diagonal elements $\PP_{\alpha\alpha}^*$ for the extended equilibrium must be chosen as
\begin{align}
    \PP_{\alpha\alpha}^{*}=\PP_{\alpha\alpha}^{\rm eq}
   + \delta t\left(\frac{2-\omega}{2\rho\omega}\right)\partial_\alpha \tilde{Q}_{\alpha\alpha\alpha},\label{eq:Paastar1}
\end{align}
where spatial derivative is evaluated using a second-order central difference scheme. 
Hence, the extended equilibrium $f_{i}^*$ is specified by using the product-form (\ref{eq:prod}),
\begin{align}
	f_{i}^*&= \rho \Psi_i(\bm{u},\bm{\PP}^*,\bm{\lambda}).
\label{eq:27eqext}
\end{align}
We shall now proceed with the derivation of the Navier--Stokes equations in the hydrodynamic limit of the proposed extended LBGK model.

\subsection{Hydrodynamic limit with extended equilibrium}
\label{sec:CE}
Taylor expansion of the shift operator in (\ref{eq:LBGK}) to second order gives,
\begin{equation}
	\left[\delta t D_i+\frac{\delta t^ 2}{2}D_iD_i\right]f_i=\omega(f_i^{*}-f_i),\label{eq:Taylor}
\end{equation}
where $D_i$ is the derivative along the characteristics,
\begin{equation}
	D_i=\partial_t + \bm{v}_i\cdot\nabla.
\end{equation}
Introducing the multi-scale expansion,
\begin{align}
	f_i&= f_i^{(0)} + \delta t f_i^{(1)} + \delta t^2 f_i^{(2)} + O(\delta t^3),  \\
	f_i^* &= f_i^{*(0)} + \delta t f_i^{*(1)} + \delta t^2 f_i^{*(2)} + O(\delta t^3), \\
	\partial_t &= \partial_t^{(1)} + \delta t\partial _t^{(2)} + O(\delta t^2), 
\end{align}
substituting into (\ref{eq:Taylor}) and using the notation,
\begin{align}
	&D_i^{(1)}=\partial_t^{(1)}+\bm{v}_i\cdot \nabla,
\end{align}
we obtain, from zeroth through second order in the time step $\delta t$,
\begin{align}
	&f_i^{(0)}=f_i^{*(0)}=f_i^{\rm eq},\label{eq:zeroCE}\\
	&D_i^{(1)}f_i^{(0)}=-\omega f_i^{(1)}+\omega f_i^{*(1)},\label{eq:oneCE}\\
	&\partial_t^{(2)}f_i^{(0)}+\bm{v}_i\cdot \nabla f_i^{(1)} -\frac{\omega}{2}D_i^{(1)} f_i^{(1)}+\frac{\omega}{2}D_i^{(1)} f_i^{*(1)}\nonumber\\
	&=-\omega f_i^{(2)}+\omega f_i^{*(2)}.\label{eq:twoCE}
\end{align}

With (\ref{eq:zeroCE}), the mass and the momentum conservation (\ref{eq:density}) and (\ref{eq:momentum}) imply the solvability conditions,
\begin{align}
	&\sum_{i=0}^{Q-1} f_i^{*(k)} = \sum_{i=0}^{Q-1} f_i^{(k)}=0,\ k=1,2\dots; \label{eq:density_orders} \\
	& \sum_{i=0}^{Q-1} \bm{v}_i f_i^{*(k)} = \sum_{i=0}^{Q-1} \bm{v}_i f_i^{(k)}=0,\ k=1,2,\dots.\label{eq:momentum_orders}
\end{align}

With the equilibrium (\ref{eq:27eq}), taking into account the solvability condition (\ref{eq:density_orders}) and (\ref{eq:momentum_orders}), and also making use of the equilibrium pressure tensor (\ref{eqn:Peq}) and (\ref{eqn:PMB}), the first-order equation (\ref{eq:oneCE}) implies the following relations for the density and the momentum,
\begin{align}
	&\partial_t^{(1)}\rho=-\nabla\cdot (\rho\bm{u}),\label{eq:density1}\\
	&\partial_t^{(1)}(\rho\bm{u})=
	-\nabla\cdot(p\bm{I}+\rho \bm{u}\otimes\bm{u}).\label{eq:momentum1}
\end{align}
Moreover, the first-order constitutive relation for the nonequilibrium pressure tensor $\bm{P}^{(1)}$ is found from (\ref{eq:oneCE}) as follows,
\begin{align}
	&-\omega \bm{P}^{(1)}+\omega \bm{P}^{*(1)}=\partial_t^{(1)}\bm{P}^{\rm eq}+\nabla \cdot \bm{Q}^{\rm eq},\label{eq:pressure1}
\end{align}
where
\begin{align}
	&\bm{P}^{(1)}=\sum_{i=0}^{Q-1} \bm{v}_i\otimes \bm{v}_i f_i^{(1)},\\
	&\bm{P}^{*(1)}=\sum_{i=0}^{Q-1} \bm{v}_i\otimes \bm{v}_i f_i^{*(1)}.
\end{align}
With the help of (\ref{eqn:PMB}), (\ref{eqn:Qeq}) and (\ref{eqn:QMB}), the first-order constitutive relation (\ref{eq:pressure1}) is transformed to make explicit the contribution of the anomalous term (\ref{eqn:Deviation}),
\begin{align}
	-\omega \bm{P}^{(1)}+\omega \bm{P}^{*(1)}=\nabla\cdot\tilde{\bm{Q}}+\left(\partial_t^{(1)}\bm{P}^{\rm MB}+\nabla\cdot \bm{Q}^{\rm MB}\right).
	\label{eq:const1}
\end{align}
The last term is evaluated using (\ref{eq:density1}) and (\ref{eq:momentum1}) to give,
\begin{equation}
	\partial_t^{(1)}\bm{P}^{\rm MB}+\nabla\cdot \bm{Q}^{\rm MB}=\rho RT\left(\nabla \bm{u}+\nabla\bm{u}^\dagger\right),\label{eq:rate}
\end{equation}
where $(\cdot)^\dagger$ denotes transposition. Combining (\ref{eq:rate}) and (\ref{eq:const1}), the first-order constitutive relation becomes,
\begin{align}
	-\omega \bm{P}^{(1)}=\left(\nabla\cdot\tilde{\bm{Q}}-\omega \bm{P}^{*(1)}\right)+\rho RT\left(\nabla \bm{u}+\nabla\bm{u}^\dagger\right).
	\label{eq:const12}
\end{align}
Note that, if we would have used the equilibrium $f_i^{\rm eq}$ instead of the extended equilibrium $f_i^*$ in (\ref{eq:LBGK}), at this stage of the derivation we get instead of (\ref{eq:const12}),
\begin{align*}
	-\omega \bm{P}^{(1)}=\nabla\cdot\tilde{\bm{Q}}+\rho RT\left(\nabla \bm{u}+\nabla\bm{u}^\dagger\right).
\end{align*}
The anomalous term $\nabla\cdot\tilde{\bm{Q}}$ cannot be canceled in the latter expression, rather, by choosing $T=T_L$ (\ref{eq:TL}), its effect can be ignored but only under the assumption of an asymptotically vanishing flow velocity. 
In contrast, using the present formulation, the cancellation is possible by finding the corresponding expression for the correction term $\bm{P}^{*(1)}$, to which end we need to consider the second-order contribution to the momentum equation. Applying the solvability condition (\ref{eq:density_orders}) and (\ref{eq:momentum_orders}) to the second-order equation (\ref{eq:twoCE}), we obtain,
\begin{align}
	&\partial_t^{(2)}\rho=0,\label{eq:rho2}\\
	&\partial_t^{(2)}(\rho\bm{u})=-\nabla\cdot \left[\left(1-\frac{\omega}{2}\right)\bm{P}^{(1)}+\frac{\omega}{2}\bm{P}^{*(1)}
\right].\label{eq:u2}
\end{align}
The latter is transformed by virtue of (\ref{eq:const12}),
\begin{align}
	\partial_t^{(2)} (\rho \bm{u})=&-\nabla\cdot\left[-\left(\frac{1}{\omega}-\frac{1}{2}\right)\rho RT(\nabla \bm{u}+\nabla\bm{u}^\dagger)\right]\nonumber\\
	&+\nabla\cdot\left[\left(\frac{1}{\omega}-\frac{1}{2}\right)\nabla\cdot \tilde{\bm{Q}}-
	\bm{P}^{*(1)}\right].
	\label{eq:momentum2}
\end{align}
The last (anomalous) term is canceled out by choosing,
\begin{align}
	\bm{P}^{*(1)}=\left(\frac{2-\omega}{2\omega}\right)\nabla\cdot\tilde{\bm{Q}}.\label{eq:cancelation}
\end{align}
Combining the result (\ref{eq:cancelation}) with the zeroth-order (equilibrium) value, we arrive at the extended pressure tensor
\begin{align}
	\bm{P}^*&=\bm{P}^{\rm eq}+\delta t \bm{P}^{*(1)}\nonumber\\
	        &=p\bm{I}+\rho \bm{u}\otimes\bm{u}+\delta t \left(\frac{2-\omega}{2\omega}\right)\nabla\cdot\tilde{\bm{Q}}.\label{eq:Pstar}
\end{align}
Since the anomalous contribution is a diagonal tensor, cf.\ Eq.\ (\ref{eqn:Deviation}), the result (\ref{eq:Pstar}) is implemented with the extended equilibrium in the product-form by choosing the the normalized (at unit density) diagonal elements of the pressure tensor as follows,

	\begin{align}
		\PP_{\alpha\alpha}^{*}&=RT+u_{\alpha}^2+\delta t \left(\frac{2-\omega}{2\rho\omega}\right)\partial_{\alpha}\left(\rho u_{\alpha}\left(\lambda_{\alpha}^2-3 RT- u_{\alpha}^2\right)\right),\label{eq:Paastar}
	\end{align}
which is equivalent to (\ref{eq:Paastar1}).
Finally, combining the first- and second-order contributions to the density and the momentum equation, (\ref{eq:density1}), (\ref{eq:momentum1}), (\ref{eq:rho2}) and (\ref{eq:momentum2}), using a notation, $\partial_t=\partial_t^{(1)}+\delta t\partial_t^{(2)}$, and also taking into account the cancellation of the anomalous term in (\ref{eq:momentum2}), we arrive at the continuity and the flow equations as follows,
\begin{align}
	&\partial_t\rho+\nabla\cdot(\rho\bm{u})=0,\label{eq:continuity}\\
	&\partial_t \bm{u} +\bm{u}\cdot\nabla\bm{u}+\frac{1}{\rho}\nabla p+\frac{1}{\rho}\nabla\cdot \bm{\Pi}
	=0,
	\label{eq:flow}
\end{align}
where $p$ is the pressure of ideal gas at constant temperature $T$,
	\begin{align}
		p=\rho RT,
	\end{align}
	$\bm{\Pi}$ is the viscous pressure tensor,
\begin{align}
	\bm{\Pi}=-\mu\bm{S},\label{eq:visc}
\end{align}
with $\bm{S}$ the rate of strain,
\begin{align}
	\bm{S}=\nabla \bm{u}+\nabla\bm{u}^\dagger,
	\label{eq:strain}
\end{align}
and $\mu$ the dynamic viscosity,
\begin{align}
	\mu=\left(\frac{1}{\omega}-\frac{1}{2}\right)p\delta t.
\end{align}

\section{Numerical results} \label{Sec:NUMRES}

The above considerations can be summarized as follows: Because of the third-order moment anomaly (\ref{eqn:Deviation}), the LBGK equation (\ref{eq:LBGK}) with the product-form equilibrium (\ref{eq:27eq}) is restricted in several ways, namely:

\noindent (i) The temperature is restricted to a single value, the lattice reference temperature $T_L$ (\ref{eq:TL});
	
\noindent (ii) The flow velocity has to be asymptotically vanishing;
	
\noindent (iii) Stretched velocities amplify these restrictions by making it impossible to cancel even the linear (in velocity) anomaly in all the directions simultaneously.

Note that, in addition to all of the restrictions above, when using the conventional second-order equilibrium obtained by retaining the terms up to the order of $\sim u_{\alpha}u_{\beta}$ in (\ref{eq:27eq}), the anomaly becomes not only confined 
to the diagonal elements $Q^{\rm eq}_{\alpha\alpha\alpha}$ but 
also contaminates the off-diagonal elements $Q^{\rm eq}_{\alpha\beta\beta}$.
While the diagonal anomaly (\ref{eqn:Deviation}) is genuine, that is, it is caused by the geometry of the discrete velocities, this additional off-diagonal deviation is due to an unsolicited second-order truncation of the product-form equilibrium (\ref{eq:27eq}).

The proposed revision of the LBGK model is based on extending the product-form equilibrium such that the anomaly of the diagonal third-order moment is compensated in the hydrodynamic limit by counter terms, which are added to the diagonal of the equilibrium pressure tensor. With this, all three restrictions mentioned above are addressed at once, without making a special distinction between the temperature, flow velocity or stretching as separate causes for the anisotropy.

In  this  section, we shall access  accuracy and performance of the proposed LB model in a variety of scenarios of activating spurious anisotropy. 
First, we test Galilean invariance, isotropy and temperature independence of the model with both regular and rectangular lattices in the simulation of a decaying shear wave.
Second, we validate the model for the more complex case of decaying homogeneous isotropic turbulence and show the effectiveness of using higher temperatures in saving compute time. Third, we investigate the applicability of the proposed model with stretched lattices in a periodic double shear layer flow, in a laminar flow over a flat plate, and finally in the case of the turbulent channel flow. In the simulations below, the gas constant was set to $R=1$ and Grad's approximation  was employed for the wall boundary condition as proposed in \citep{dorschner2015grad}.

\subsection{Galilean invariance, isotropy and temperature independence test}
To probe the Galilean invariance and temperature independence of the model, the numerical kinematic viscosity $\nu = \mu / \rho$ (\ref{eq:nu}) is measured for the decay of a plane shear wave.
First, we consider the axis-aligned setup, with the initial condition,
	\begin{equation} 
	\rho  = \rho _0,\  u_x = a_0\sin (2\pi y/L_y),\  u_y = {\rm Ma}\sqrt{T}, 
	\end{equation}
\begin{figure}[h]
		\centering
		\includegraphics[width=0.5\textwidth]{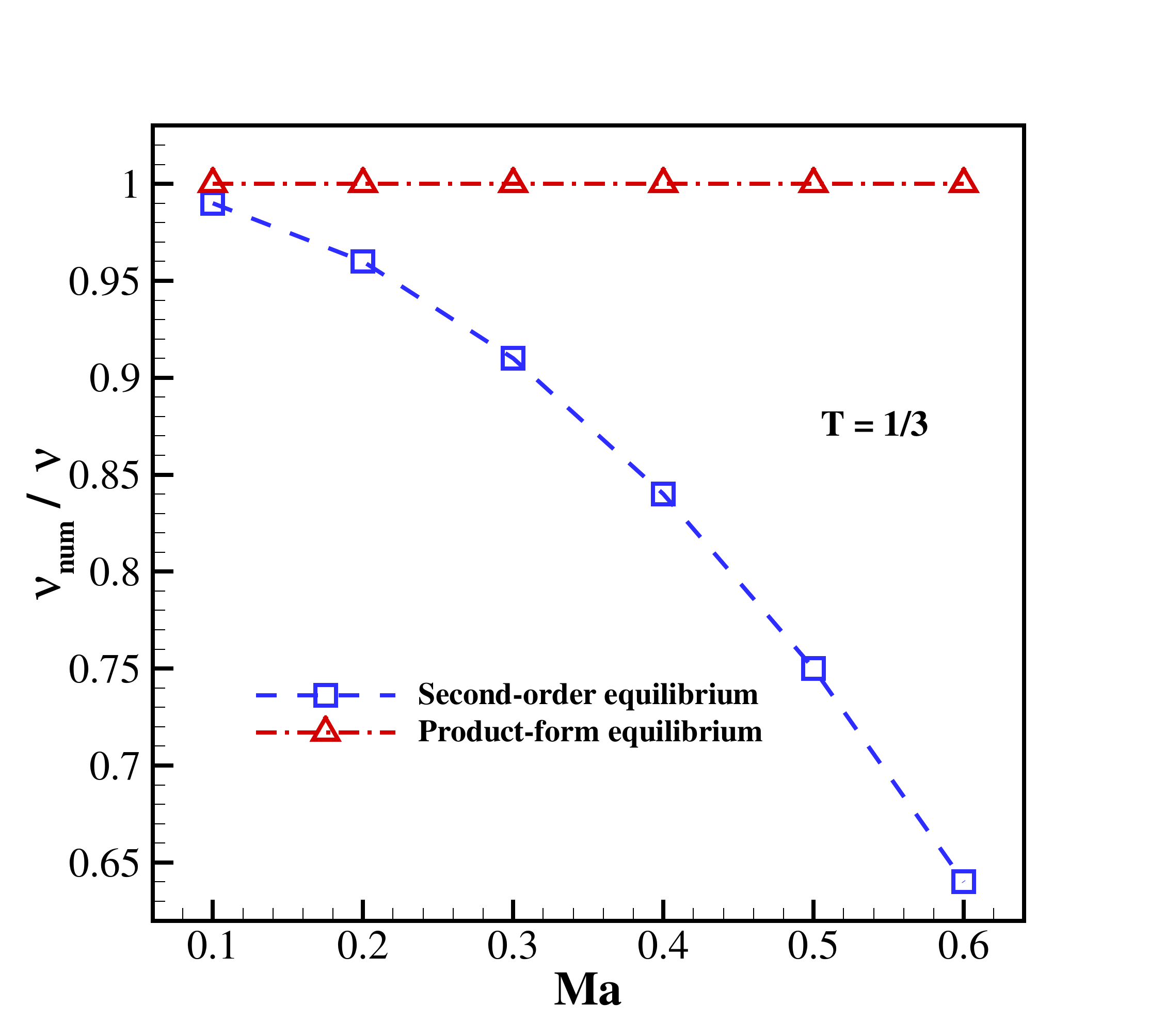}
		\caption{Numerical measurement of viscosity for axis-aligned setup at temperature $T=1/3$ for different velocities.}
		\label{fig:visc_product1}
\end{figure}
\begin{figure}[h]
		\centering
		\includegraphics[width=0.5\textwidth]{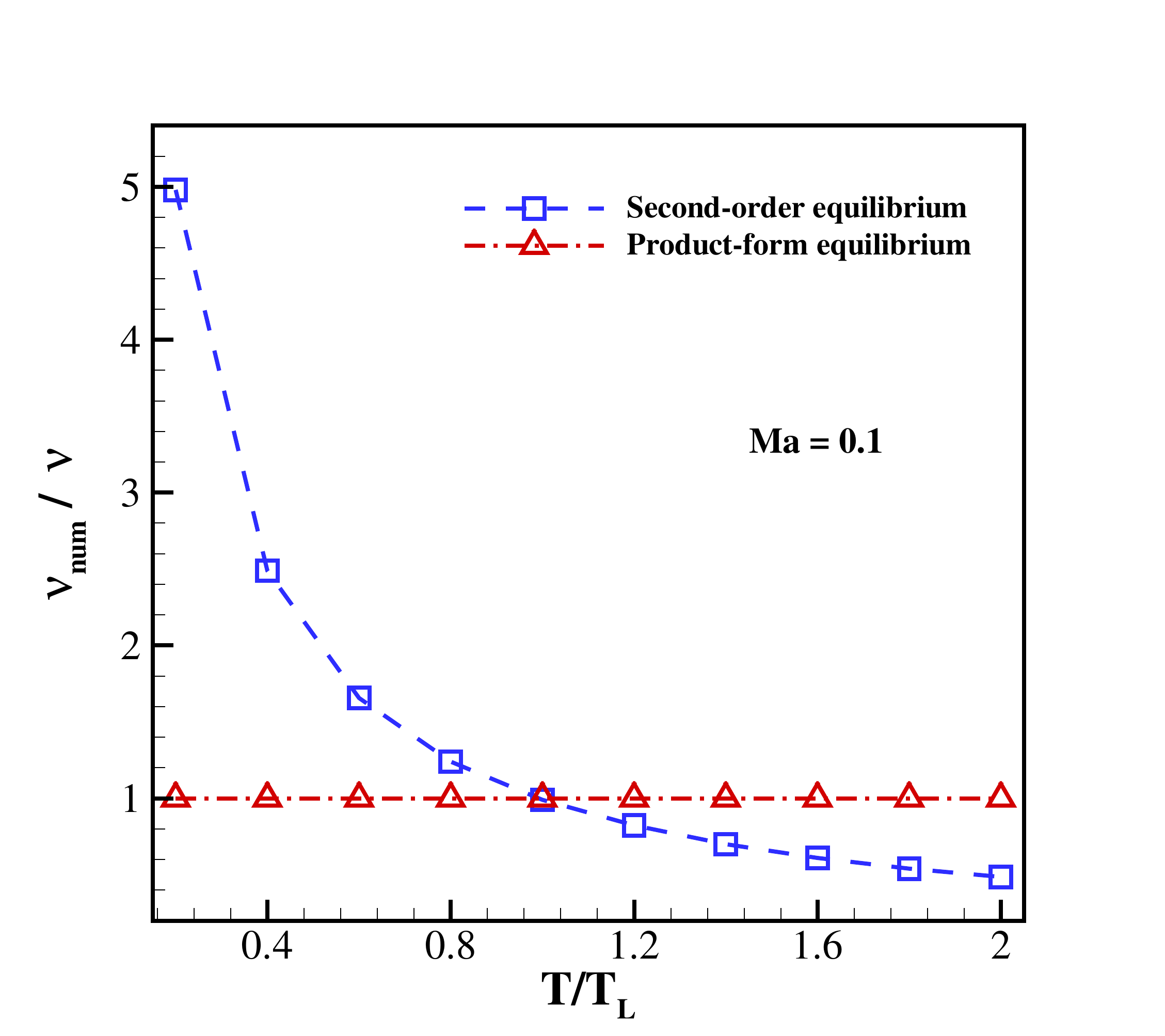}
		\caption{Numerical measurement of viscosity for axis-aligned setup at Mach number $Ma=0.1$ for different temperatures.}
		\label{fig:visc_product2}
\end{figure}
where  ${\rm Ma}=u_0/\sqrt{T}$ is the advection Mach number, $a_0 = 0.001$ is the amplitude, $L_y = 200$ is number of grid nodes in the $y$ direction, $\rho_0 = 1$.
Periodic boundary conditions are imposed in both $x$- and $y$- directions. The viscosity is measured by fitting an exponential to the time decay of maximum flow velocity $u_x$.
In this special case, the diagonal anomaly (\ref{eqn:Deviation}) is dormant and does not trigger any spurious effects because the derivatives $\partial_x\tilde{Q}_{xxx}$ and $\partial_y\tilde{Q}_{yyy}$ both vanish. Consequently, the extended equilibrium (\ref{eq:27eqext}) becomes equivalent to the product-form equilibrium (\ref{eq:27eq}) in this case.

In order to compare with the standard LBGK, the standard velocities $\lambda_{\alpha}=1$ were used in this simulation. Fig.\ \ref{fig:visc_product1} and Fig.\ \ref{fig:visc_product2} show the importance of using the product-form equilibrium (\ref{eq:27eq}) as opposed to the conventional LBGK model with the second-order equilibrium. A strong dependence of the viscosity on the reference frame  for the second-order equilibrium can be seen in Fig. \ref{fig:visc_product1}, where the viscosity drops with increasing advection Mach number. This well-known artifact of the second-order equilibrium is due to the non-vanishing anomaly in the off-diagonal moments $Q^{\rm eq}_{\alpha\beta\beta}$, and, unlike the diagonal anomaly, is caused only by the approximate treatment of the product-form equilibrium. Moreover, as shown in Fig.\ \ref{fig:visc_product2}, even at a small enough velocity this spurious feature improves only at the lattice reference temperature $T_L$.
In contrast, as is shown in Fig.\ \ref{fig:visc_product1} and Fig.\ \ref{fig:visc_product2}, the product-form equilibrium of the present model is able to accurately predict the viscosity in this setup for a wide range of temperatures and reference frame velocities.
\begin{figure}[h]
		\centering
		\includegraphics[width=0.5\textwidth]{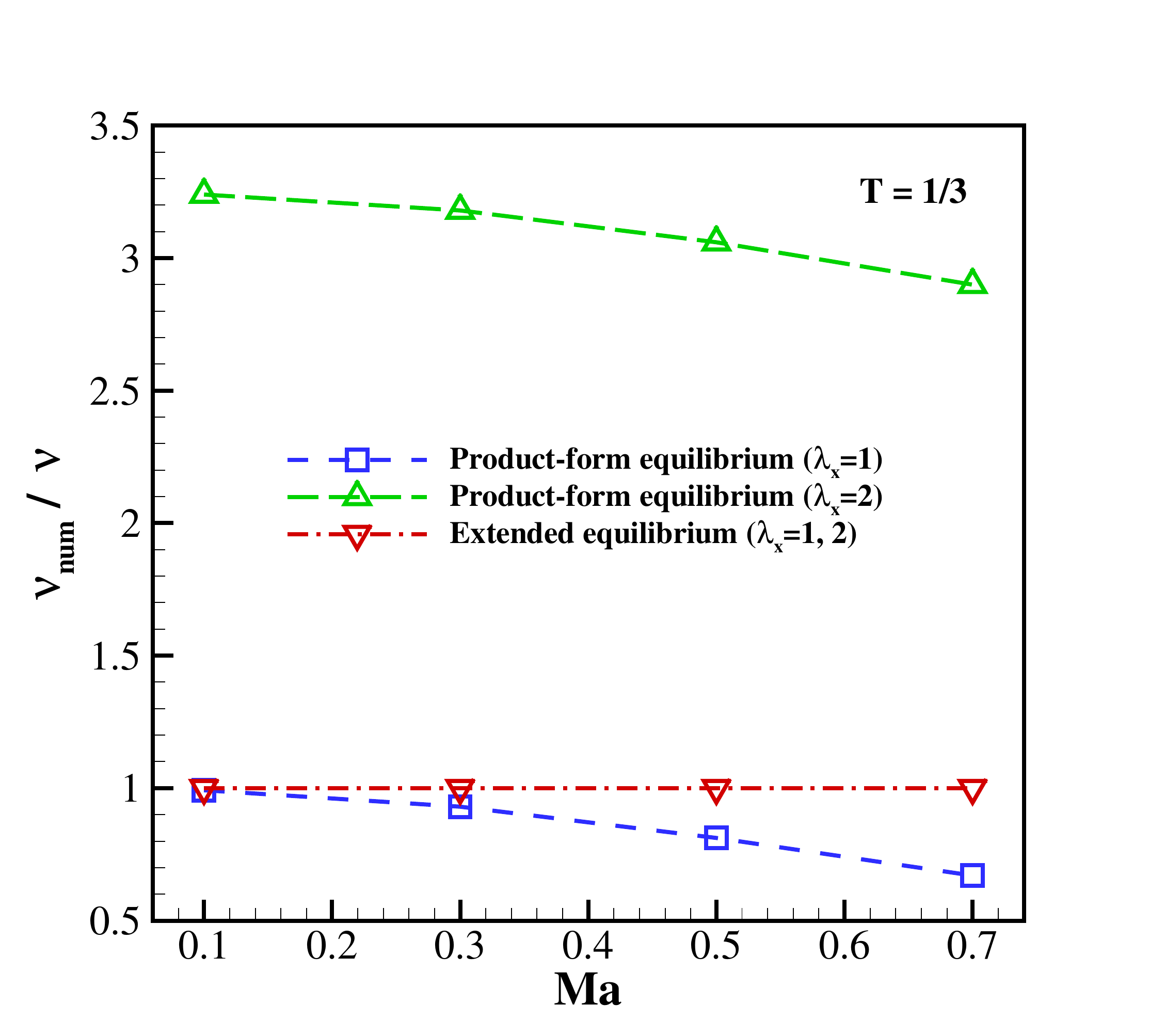}
		\caption{Numerical measurement of viscosity for rotated setup at temperature $T=1/3$ for different velocities and stretching rations.
		       }
		\label{fig:visc_rotate}
\end{figure}

Next, in order to trigger the anisotropy of the deviation terms (\ref{eqn:Deviation}) and to show the necessity of using the extended equilibrium, the shear wave is rotated by $\pi/4$. The anisotropy is further increased by also conducting simulations on a stretched grid with $\lambda_x=2$. The temperature is kept at $T=1/3$. The viscosity measurement is shown in Fig.\ \ref{fig:visc_rotate} for different advection Mach numbers and stretching factors. It can be observed that the model lacks Galilean invariance for larger velocities when using the product-form equilibrium without correction (\ref{eq:27eq}). Furthermore, the stretching factor $\lambda_x=2$ results in a significant hyper-viscosity since the deviation (\ref{eqn:Deviation}) in this case amounts to a large positive number. However, once the correction term is included and the extended equilibrium (\ref{eq:27eqext}) is used, the present model recovers the imposed viscosity, independent of the frame velocity and stretching factor.

\subsection{Decaying homogeneous isotropic turbulence}\label{sec:HIT}
In order to further validate the model as a reliable method for the simulation of complex flows and to show the application of using higher temperatures, decaying homogeneous isotropic turbulence was considered. The initial condition, in a box of the size $L \times L \times L$, was set at unit density and constant temperature along with a divergence-free velocity field, which follows the specified energy spectrum,
\begin{align}
      E(\kappa) &= A \kappa^4 exp\left(-2(\kappa/\kappa_0)^2\right),
\end{align}
where $\kappa$ is the wave number, $\kappa_0$ is the wave number at which the spectrum peaks and $A$ is the parameter that controls the initial kinetic energy \citep{samtaney2001direct}. 
The initial velocity field is generated using a kinematic simulation as proposed in \citep{ducasse2009inertial}.
The turbulent Mach number is defined as
\begin{align*}
 Ma_t = \frac{\sqrt{\overbar{\bm{u}\cdot\bm{u}}}}{{c_s}},
\end{align*}
 where $c_s=\sqrt{T}$ is the speed of sound. The  Reynolds number is based on the Taylor microscale,
 \begin{align}
   \lambda = \frac{\overbar{u_x^2}}{\overbar{(\partial_x u_x)^2}}, \label{eq:Taylor Lamda}
 \end{align}
and is given by
 \begin{align*}
     Re_\lambda = \frac{\overbar{\rho} u_{rms}\lambda}{\mu},
 \end{align*}
where $u_{rms} = \sqrt{{\overbar{\bm{u}\cdot\bm{u}}}/{3}}$ is the root mean square (rms) of the velocity and overbar denotes the volume average over the entire computational domain.

\begin{figure}[h]
		\centering
		\includegraphics[width=0.5\textwidth]{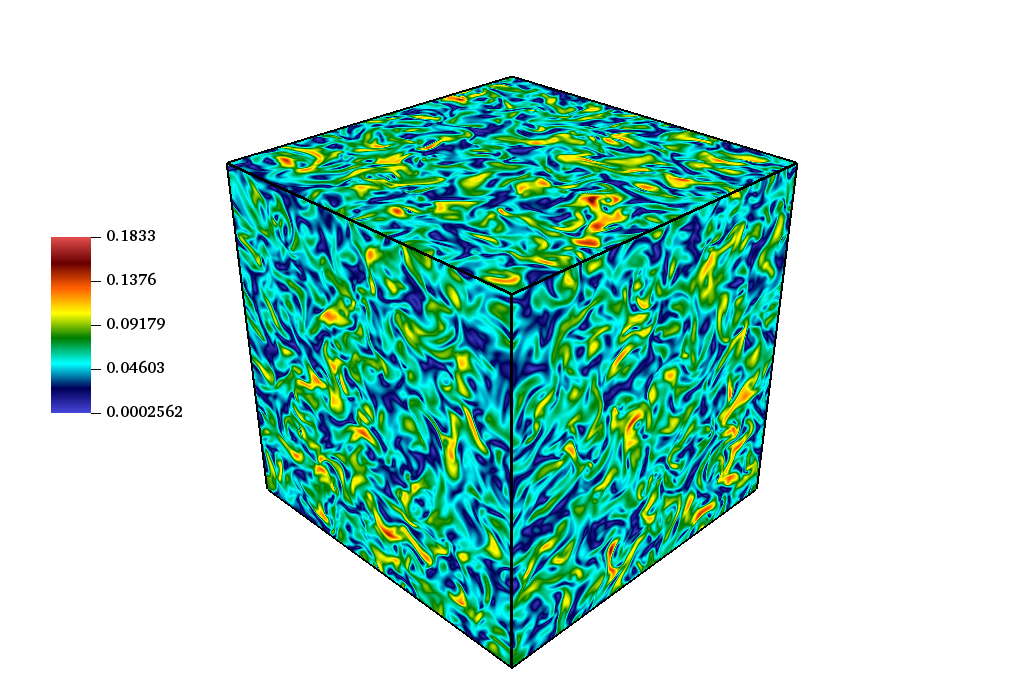}
		\caption{Velocity magnitude of the decaying homogeneous isotropic turbulence at $Ma_t=0.1$, $Re_\lambda=72$ and $t^*=1.0$ with temperature $T=0.55$. }
		\label{fig:DecayTurb}
\end{figure}

Simulations were performed at $Ma_t=0.1$, $Re_\lambda=72$, $\kappa_0=16\pi/L$, at two different temperatures, $T=1/3$ and $T=0.55$, and with $L=256$ grid points. 
Fig.\ \ref{fig:DecayTurb} shows a snapshot of the velocity magnitude at time $t^*=t/\tau = 1.0$, where $\tau=L_I/u_{rms,0}$ is the  eddy turnover time, which is defined based on the initial rms of the velocity and the integral length scale $L_I = {\sqrt{2\pi}}/{\kappa_0}$.
\begin{figure}[h]
		\centering
		\includegraphics[width=0.5\textwidth]{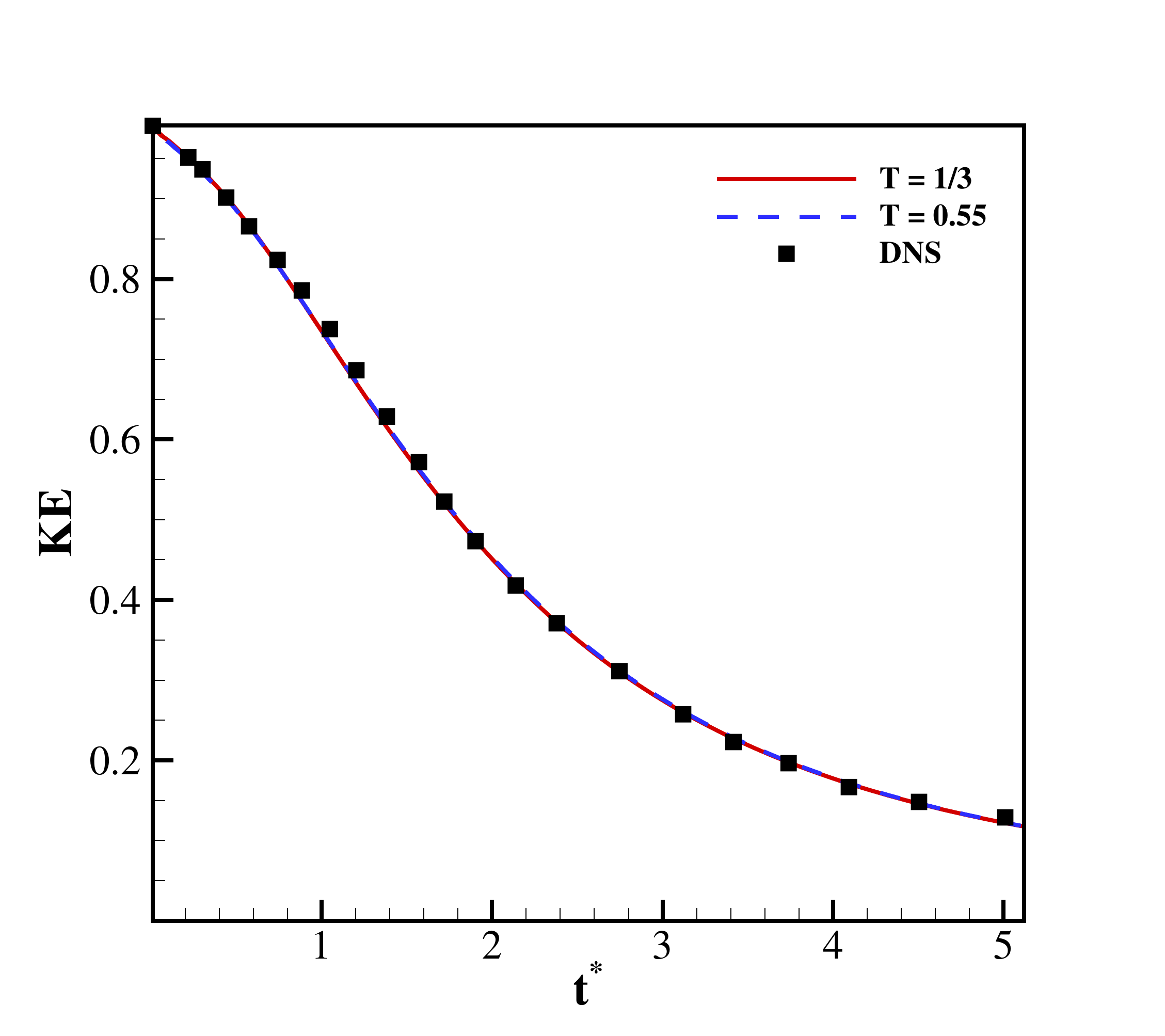}
		\caption{Time evolution of the turbulent kinetic energy for decaying isotropic turbulence at $Ma_t=0.1$, $Re_\lambda=72$.
		Lines: present model; symbol: DNS \cite{samtaney2001direct}. }
		\label{fig:Decay_KE}
\end{figure}
\begin{figure}[h]
		\centering
		\includegraphics[width=0.5\textwidth]{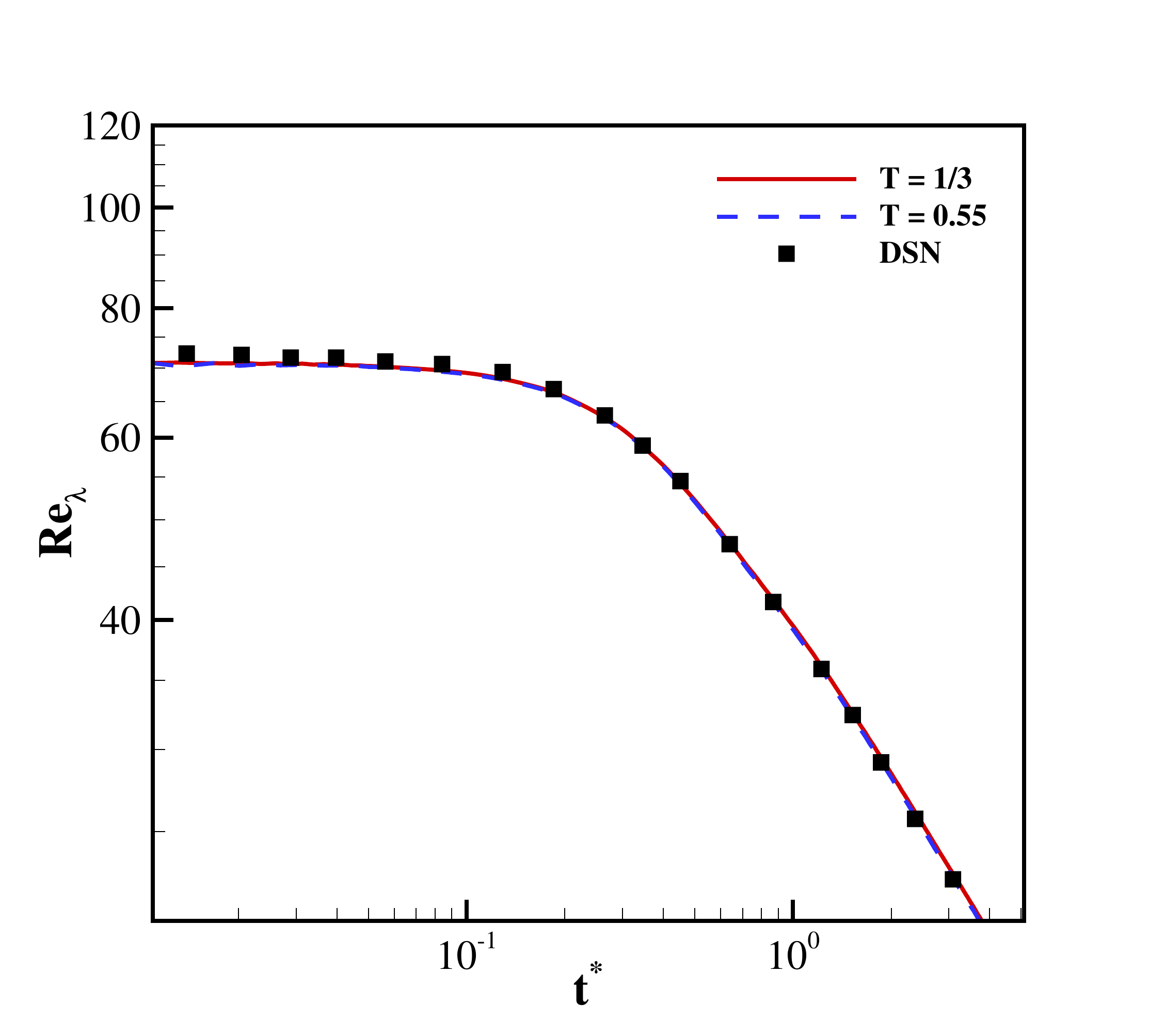}
		\caption{Time evolution of the Taylor microscale Reynolds number for decaying isotropic turbulence at $Ma_t=0.1$, $Re_\lambda=72$. 
		Lines: present model; symbol: DNS \cite{samtaney2001direct}}
		\label{fig:Decay_RE}
\end{figure}
To quantitatively assess the accuracy of the model at different temperatures, the time evolution of the turbulent kinetic energy,
\begin{align*}
     KE = \frac{1}{2} \overbar{\bm{u}\cdot\bm{u}},
\end{align*}
and of the Taylor microscale Reynolds number are compared in Fig.\ \ref{fig:Decay_KE} and Fig.\ \ref{fig:Decay_RE} with results from direct numerical simulations (DNS) \citep{samtaney2001direct}. 
It is apparent that the two working temperatures yields almost identical results that agree well with the DNS simulation.
This indicates that the correction terms do not degrade the accuracy of the model at higher temperatures, even though the magnitude of error term (\ref{eqn:Deviation}) is higher due to amplification of the linear term.  

The immediate advantage of using the present model at a temperature higher than the lattice temperature $T_L=1/3$ is that it effectively increases the characteristic velocity (here $u_{rms,0}$) and therefore the time step by a factor of $\sqrt{T/T_L}$. 
A larger time step is equivalent to fewer number of time steps. 
The present model, therefore, speeds up the simulation by a factor of $\sqrt{T/T_L}$ compared to the conventional LBM, which can operate only at the lattice temperature $T_L$. Furthermore, this speedup strategy can be used for both steady and unsteady flows. This is in contrast to the preconditioned LBM \citep{guo2004preconditioned}, which works by altering the effective Mach number and therefore reduces the disparity between the speeds of the acoustic wave propagation and the waves propagating with the fluid velocity, cf.\ \citep{guo2004preconditioned}. This makes preconditioned LBM restricted to steady state applications. In contrast, the present model enables us to increase the speed of sound without changing the Mach number. This increases the effective time step of the solver. Therefore, the present model enhances the computational efficiency by decreasing the number of required time steps.

\subsection{Periodic double shear layer}
The next validation case to test the accuracy of the proposed model with the stretched lattice is the periodic double shear layer flow with the initial condition,
\begin{align*}
    u_x &= \left\{\begin{matrix}
    u_0 \tanh(\kappa(y/L-0.25)), & y \leq L/2, \\
    u_0 \tanh(\kappa(0.75-y/L)), & y > L/2, 
\end{matrix}\right.
\\
    u_y &= \delta u_0 \sin(2\pi(x/L+0.25)),
\end{align*}
where $L$ is the domain length in both $x$ and $y$ directions, $u_0=0.1$ is characteristic velocity, $\delta=0.05$ is a perturbation of the $y$-velocity and $\kappa=80$ controls the width of the shear layer. The Reynolds number is set to $Re = u_0 L/ \nu = 10^4$ and the temperature is $T=1/3$.

\begin{figure}[h]
		\centering
		\includegraphics[width=0.5\textwidth]{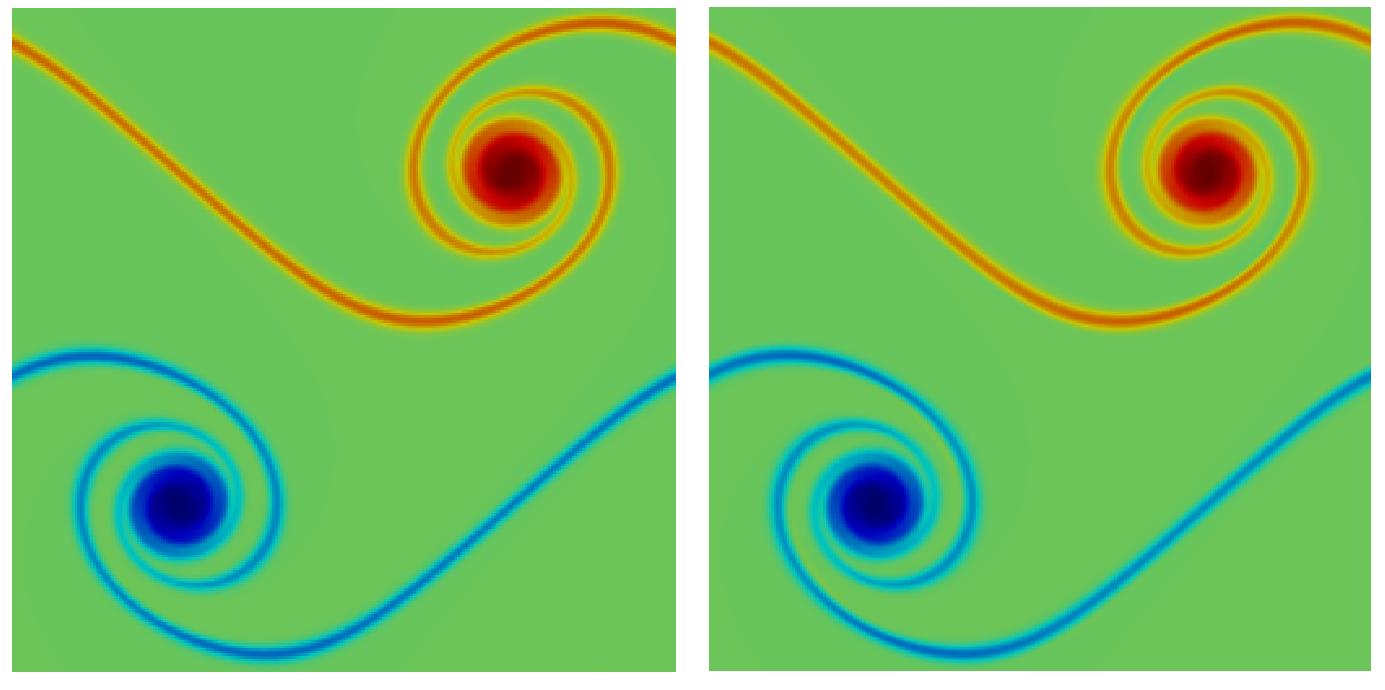}
		\caption{Vorticity field for double shear layer flow at $t^*=1$ with regular lattice (left) and stretched lattice (right). }
		\label{fig:DLF_Vor}
\end{figure}

\begin{figure}[h]
		\centering
		\includegraphics[width=0.5\textwidth]{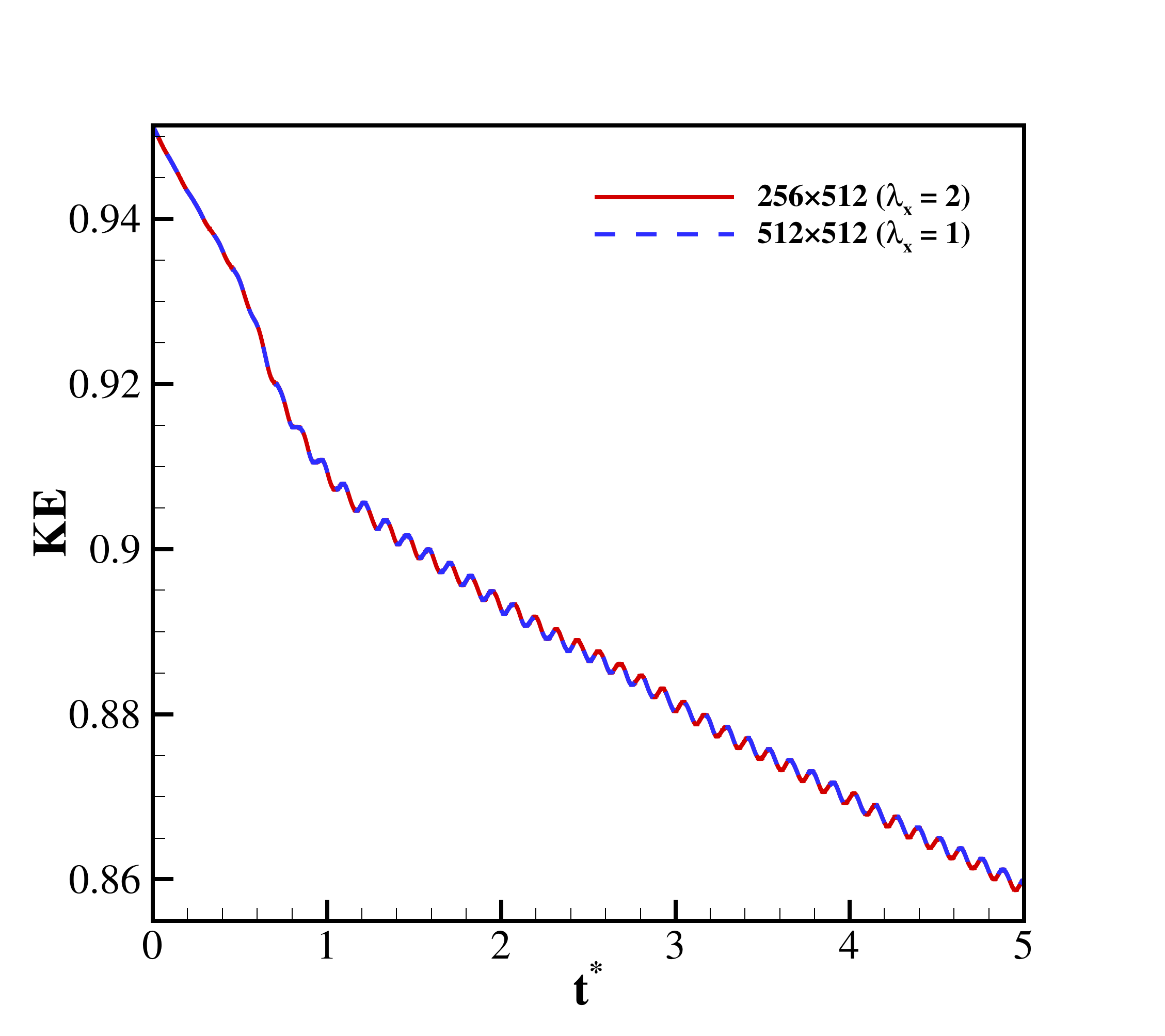}
 	    \includegraphics[width=0.5\textwidth]{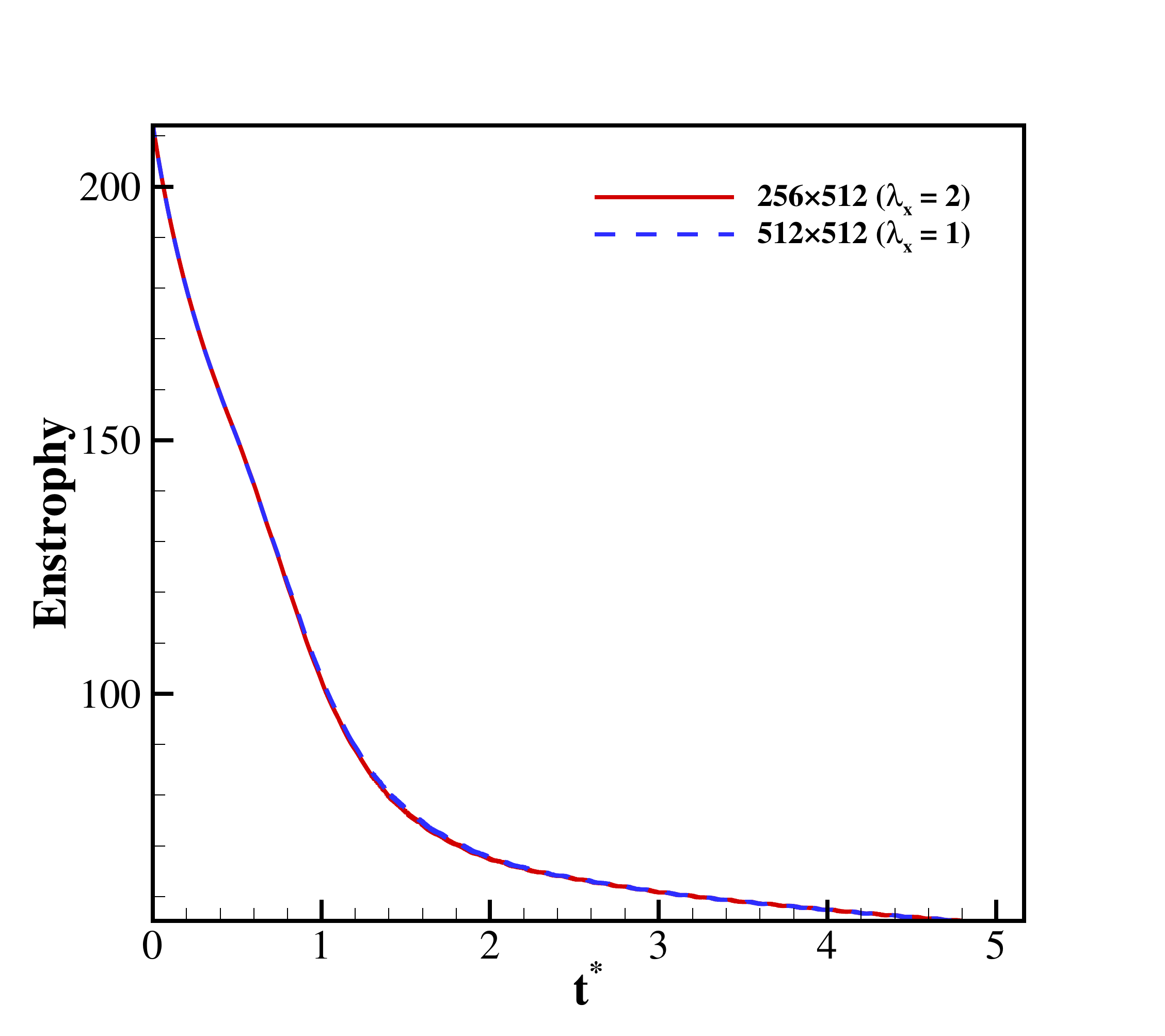}
		\caption{Evolution of kinetic energy (top) and enstrophy (bottom) for double shear layer flow at $Re=10^4$. }
		\label{fig:DLF_KENS}
\end{figure}

Fig.\ \ref{fig:DLF_Vor} shows the vorticity field at non-dimensional time $t^*=t u_0/L=1$ using the conventional square lattice $\lambda_x=\lambda_y=1$ and the rectangular lattice with $\lambda_x=2$, $\lambda_y=1$. Both lattice models perform qualitatively same.
To quantify the effect of stretching on the accuracy, the time evolution of the mean kinetic energy $KE=\overbar{u^2}/u_0^2$, and of the mean enstrophy $E=\overbar{\Omega^2}/\frac{u_0^2}{L^2}$, with $\Omega$ the vorticity magnitude, are compared in Fig.\ \ref{fig:DLF_KENS}.
The results show only minor discrepancies, which indicates the validity of the model also on stretched meshes.

\subsection{Laminar boundary layer over a flat plate}
\begin{figure}[]
		\centering
		\includegraphics[width=0.5\textwidth]{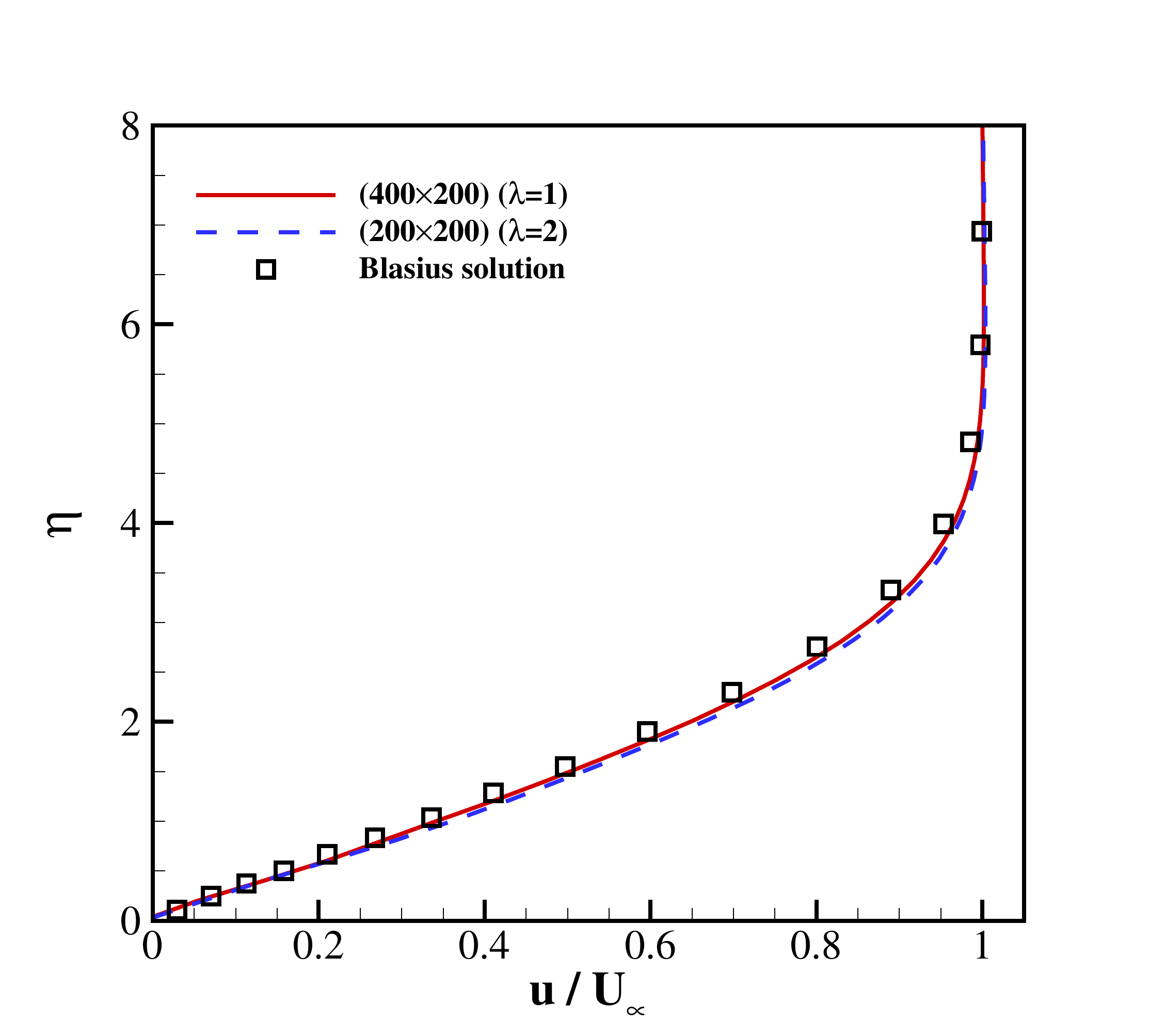}	\caption{Comparison of the velocity profile for flow over a flat plate at different stretching ratios. Lines: present model; symbols: Blasius solution. %
		}
		\label{fig:FlatVel}
\end{figure}

\begin{figure}[]
		\centering
		\includegraphics[width=0.5\textwidth]{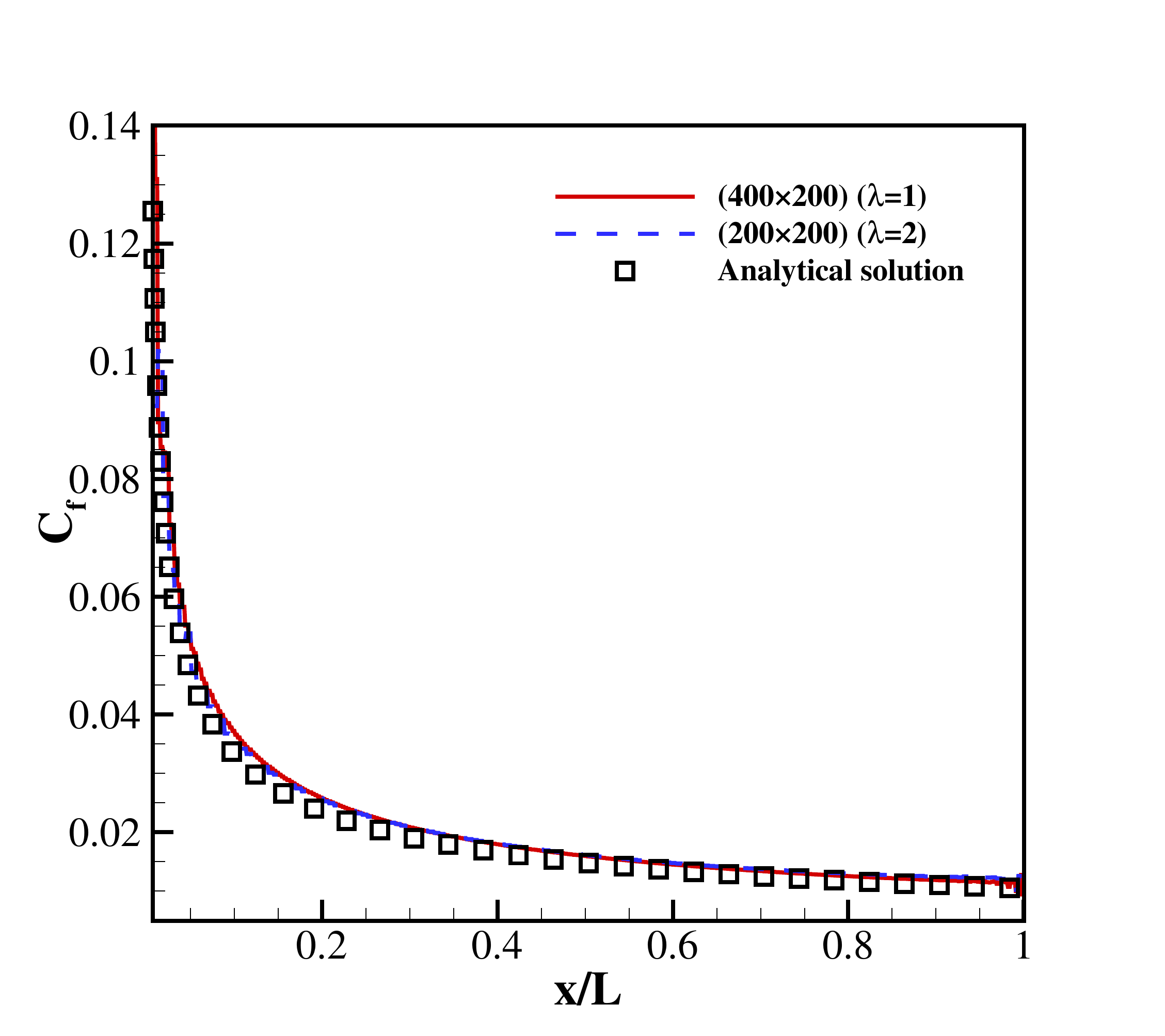}	\caption{Comparison of the skin friction coefficient for flow over a flat plate at different stretching ratio. Lines: present model; symbols: analytical solution.}
		\label{fig:FlatCf}
\end{figure}

The next test case validates our model for wall-bounded flows. We consider the laminar flow over a flat plate with an incoming Mach number $Ma_\infty=u_\infty/\sqrt{ T_\infty} =0.1$, temperature $T_\infty = 1/3$ and Reynolds number $Re = \rho_\infty u_\infty L / \mu = 4000$, where $L$ is the length of flat plate. Since the flow gradients in the transverse $y$-direction are much larger compared to the gradients in the streamwise $x$-direction, the mesh can be stretched in $x$-direction without significantly affecting the accuracy of the results.
The computational domain was set to $[L_x \times L_y]=[200 \times 200]$ and a rectangular lattice with $\lambda_x=2$ was used. 
The flat plate starts at a distance of $L_x/4$ from the inlet and symmetry boundary conditions were imposed at $0 \leq x \leq L_x/4$.
In Fig.\ \ref{fig:FlatVel}, the horizontal velocity profile at the end of the plate is compared with the results of a regular lattice and with the Blasius similarity solution, where $\eta$ is the dimensionless coordinate \citep{white1979fluid},
\begin{align*}
    \eta = y \sqrt{\frac{u_\infty}{\nu x}}.
\end{align*}
It can be seen that results for the regular and the rectangular lattice nearly coincide and agree well with the Blasius solution. 
Thus, the model achieves accurate results with half of grid points compared to the regular lattice. 
Furthermore, the distribution of skin friction coefficient over the plate, 
\begin{align*}
    C_f = \frac{\tau_{wall}}{\frac{1}{2} \rho_\infty u_\infty^2},
\end{align*}
with the wall shear stress $\tau_{wall} = \mu (\frac{\partial u}{\partial y})_{y=0}$, is shown in Fig.\ \ref{fig:FlatCf} in comparison with the analytical solution $C_f=0.664 / \sqrt{\text{Re}_x}$, where $\text{Re}_x=u_\infty x /\nu$ \citep{white1979fluid}. 
Also here, the results of the model with the regular and the stretched velocities are almost identical and in good agreement with the analytical solution.

\subsection{Turbulent channel flow}
In the final test case, we assess the accuracy and performance of the extended LBM for the turbulent flow in a rectangular channel, for which many numerical \citep{kim1987turbulence,moser1999direct,lee2015direct} and experimental \citep{eckelmann1974structure,kreplin1979behavior} results are available. The channel geometry was chosen as $[5.6H \times 2H \times 2H]$, where $H$ is the channel half-width. The friction Reynolds number,
\begin{align*}
    \text{Re}_\tau = \frac{u_\tau H}{\nu},
\end{align*}
 based on the friction velocity $u_\tau = \sqrt{\tau_w/\rho}$, was set to $Re_\tau=180$. The initial friction velocity was estimated by 
\begin{align*}
    u_\tau = \frac{u_0}{\frac{1}{\kappa} \text{ln} Re_\tau + 5.5},
\end{align*}
where $\kappa=0.41$ is the von K\'{a}rm\'{a}n constant and $u_0=0.1$ is the mean center-line velocity. Periodic boundary conditions were imposed in the streamwise $x$-direction and the spanwise $z$-direction. The flow was driven by a constant body force in the $x$-direction, 
\begin{align*}
    \mathrm{g} = \text{Re}_\tau^2 \nu^2 / H^3.
\end{align*}
In order to accelerate the transition to turbulence, a non-uniform divergence-free forcing field as proposed in \citep{wang2016lattice} was added to the flow for some period of time, until $t^*= t H/u_\tau =5$.  %

Similar to the previous test case, grid stretching in $x$-direction with $\lambda_x=1.4$ was used in order to reduce the number of grid points in that direction while the temperature was set to $T=0.55$, same as in Sec.\ \ref{sec:HIT}.
\begin{figure}[]
		\centering
		\includegraphics[width=0.5\textwidth]{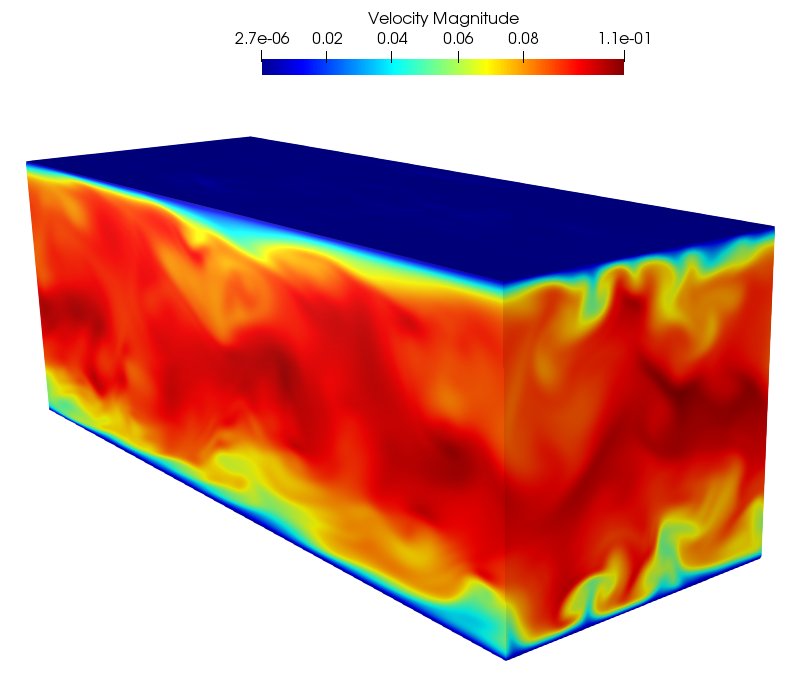}	
		\caption{Snapshot of a turbulent channel flow at $Re_\tau=180$ with $\lambda_x=1.4$.}
		\label{fig:Channel_xyz}
\end{figure}
A snapshot of the velocity magnitude is shown in Fig.\ \ref{fig:Channel_xyz}. 
\begin{figure}[]
		\centering
		\includegraphics[width=0.5\textwidth]{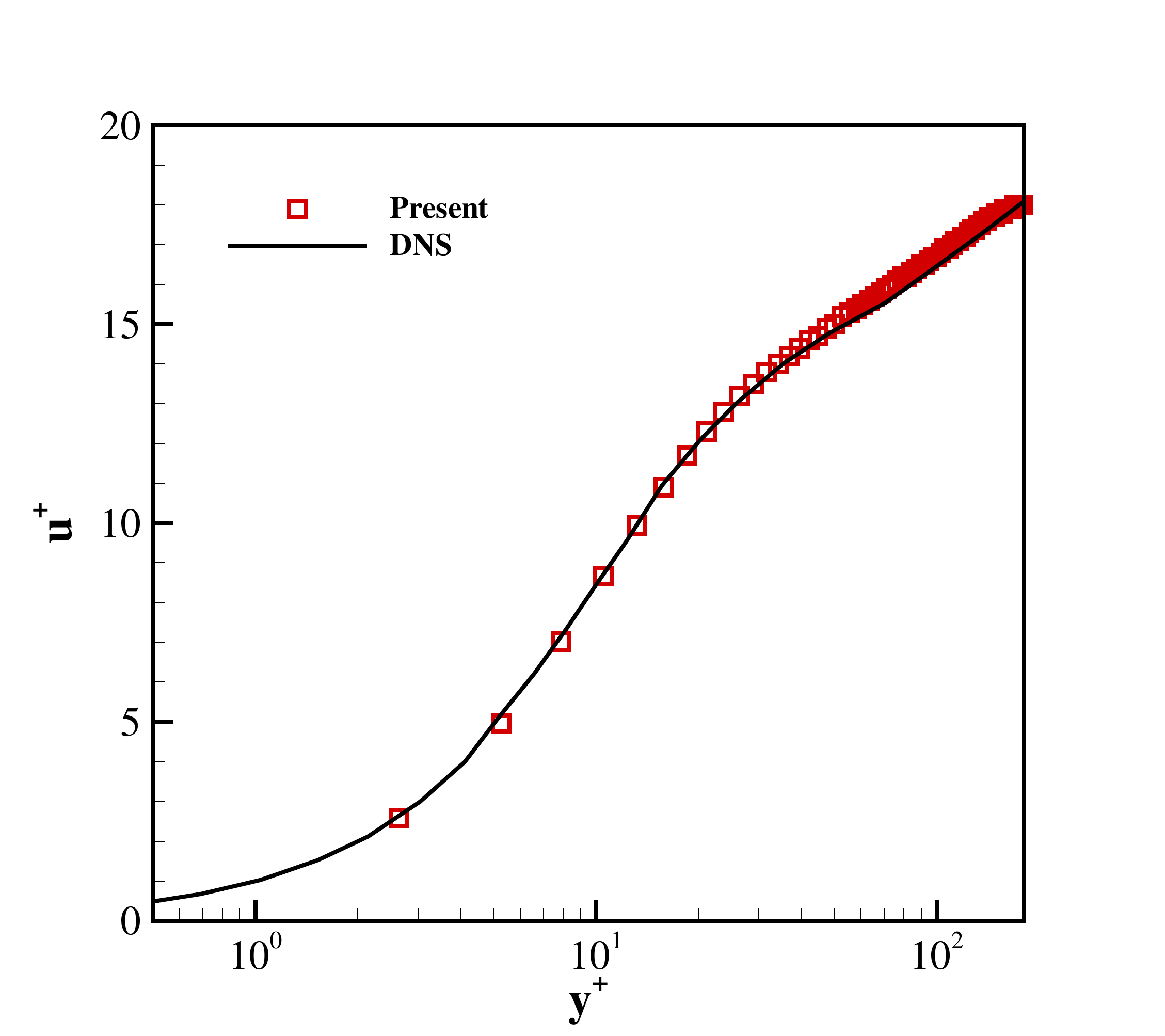}	
		\caption{Comparison of the mean velocity profile in a turbulent channel flow at $Re_\tau=180$ with $\lambda_x=1.4$.}
		\label{fig:ypup}
\end{figure}
Quantitatively, we compare the mean velocity profile with the DNS results of \citep{moser1999direct} in Fig.\ \ref{fig:ypup}.
In wall units, the mean velocity is given by $u^+ = \bar{u}/u_\tau$ and the spatial coordinate is $y^+=y u_\tau / \nu$.
The statistics are collected after $30$ eddy turnover times, i.e., after $t^* =30$. 
It is apparent that the viscous sublayer ($y^+<5$), the buffer layer ($5<y^+<30$) and the log-law region ($y^+>30$) are captured well with our 
model and the mean velocity profile agrees well with that of the DNS. 
\begin{figure}[]
		\centering
		\includegraphics[width=0.5\textwidth]{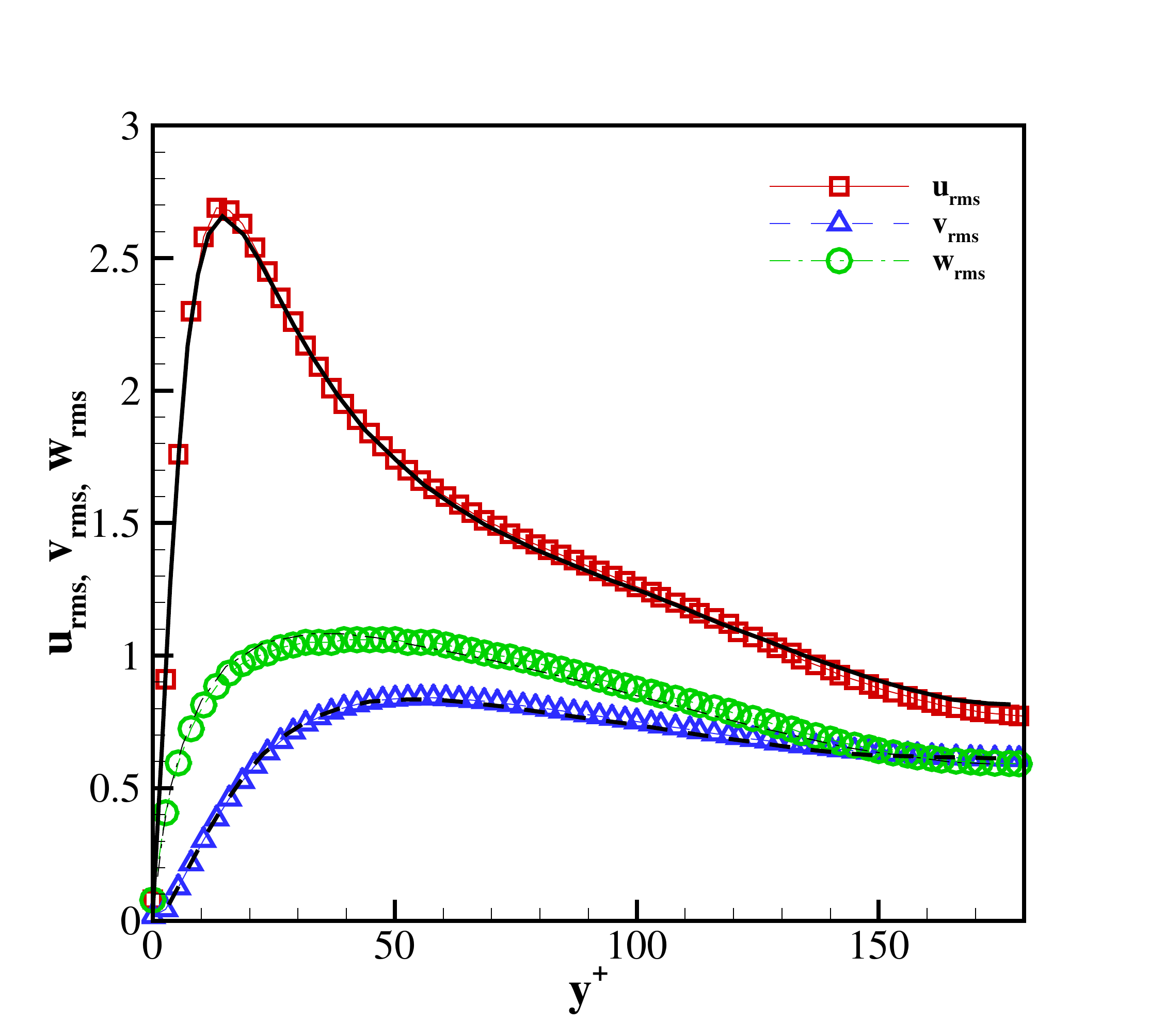}	
		\caption{Comparison of the rms of the velocity fluctuations in a turbulent channel flow at $Re_\tau=180$ with $\lambda_x=1.4$. Symbols: present model; lines: DNS \cite{moser1999direct}.}
		\label{fig:Channel_rms}
\end{figure}

For a more thorough analysis, we compare the root mean square of the velocity fluctuations with the DNS data in Fig.\ \ref{fig:Channel_rms}. Here, $u_{rms}=\sqrt{\overbar{u^{'} u^{'}}}$ and $v_{rms}$ and $w_{rms}$ are defined in a similar way. It can be seen that the results are in excellent agreement with the DNS results \citep{moser1999direct}. This demonstrates that the LBGK model, also in the presence of a severe anisotropy triggered by  stretched velocities, can be used for the simulation of high Reynolds number wall-bounded flows once the corrections are incorporated with the extended equilibrium.

\section{Conclusion} \label{Sec:Conclusion}

While even with the standard discrete speeds (\ref{eq:d3q27vel}) it is possible to develop an error-free, fully Galilean invariant kinetic model in the co-moving reference frame, it does require off-lattice particles' velocities \cite{PonD2018,PonD2020}.
Sticking with the fixed, lattice-conform velocities (\ref{eq:d3q27velSTRETCHED}), one is faced with an inevitable and persistent error, which spoils the hydrodynamic equations whenever the flow velocity is increased or the temperature deviates from the lattice reference value, or the discrete speeds are stretched differently in different directions.
We proposed an upgrade of the LBGK model to enlarge its operation domain in terms of velocity, temperature and grid stretching by suggesting an extended equilibrium.
The extended equilibrium is realized through a product-form, which allows us to compensate the diagonal third-order moment anomaly in the hydrodynamic limit by 
adding consistent correction terms to the diagonal elements of the second-order moment. 
As a result, the extended LBGK  model restores Galilean invariance and temperature independence in a sufficiently wide range, and can also be used with rectangular lattices. 
Similar to previous proposals \cite{prasianakis2007lattice,prasianakis2008lattice,prasianakis2009,saadat2019lattice,sawant2021consistent}, the relaxation term of the present model remains almost local as it uses only nearest-neighbor information for computation of the first-order derivatives in the extended equilibrium populations.
The extended LBGK model was validated in a range of benchmark problems, probing different aspects of anomaly triggered either by increased velocity or temperature deviation from the lattice reference temperature, or by grid stretching. In all cases, the extended LBGK model featured excellent performance and accuracy in both two and three dimensions. In particular, the simulation of homogeneous isotropic turbulence demonstrated the expected speed-up when a higher temperature was used, while simulations of the laminar boundary layer and of the turbulent channel flow using stretched grids demonstrated good accuracy with a reduced number of grid points.\\
Furthermore, the present model can be extended to other applications including but not limited to high-speed compressible flows, which can be  achieved by incorporating another solver for the the total energy (see, e.g., the models proposed in \citep{saadat2019lattice,sawant2021consistent}).
Advanced collision models such as multi-relaxation-time schemes can also readily be employed in the present approach, which can be beneficial when running under-resolved simulations. These two avenues shall be subject of further development and application of the extended LBM.

\section*{Acknowledgements}
This work was supported by the European Research Council (ERC) Advanced Grant 834763-PonD and the ETH Research Grant ETH-13 17-1. Computational resources at the Swiss National Super Computing Center CSCS were provided under the Grant s897. 

\bibliography{Bib}
\end{document}